\documentclass[usegraphicx]{mn2e}
\usepackage{epsfig}
\usepackage{amssymb}

\renewcommand{\H}{{\rm H}}

\newcommand{\HI}{H\,\textsc{i}\ }
\newcommand{\HII}{H\,\textsc{ii}\ }
\newcommand{\HeI}{He\,\textsc{i}\ }
\newcommand{\HeII}{He\,\textsc{ii}\ }
\newcommand{\HH}{H$_2$}
\newcommand{\Msun}{\,{\rm M_\odot}}
\newcommand{\Tvir}{T_{\rm vir}}

\newcommand{\etal}{{et al.\ }}
\newcommand{\ie}{{i.e.\ }}

\newcommand{\be}{\begin{equation}}
\newcommand{\ee}{\end{equation}}
\newcommand{\ba}{\begin{eqnarray}}
\newcommand{\ea}{\end{eqnarray}}

\title{The first miniquasar}
\author[Kuhlen \& Madau]{Michael Kuhlen \& Piero Madau\\
Department of Astronomy and Astrophysics, University of California at Santa 
Cruz, Santa Cruz, CA 95064, U.S.A.\\
(e-mail: mqk@ucolick.org, pmadau@ucolick.org)}
\date{}

\pagerange{\pageref{firstpage}--\pageref{lastpage}} \pubyear{2005}

\begin{document}

\label{firstpage}

\maketitle

\begin{abstract}
We investigate the environmental impact of the first active galactic
nuclei that may have formed $\sim150$ Myr after the big bang in
low-mass $\sim 10^6\,\Msun$ minihaloes. Using \textsc{enzo}, an adaptive-mesh
refinement cosmological hydrodynamics code, we carry out
three-dimensional simulations of the radiative feedback from
`miniquasars' powered by intermediate-mass black holes. We follow the
non-equilibrium multispecies chemistry of primordial gas in the
presence of a point source of X-ray radiation, which starts shining in
a rare high-$\sigma$ peak at $z=21$ and emits a power-law spectrum in
the 0.2-10 keV range. We find that, after one Salpeter time-scale, the
miniquasar has heated up the simulation box to a volume-averaged
temperature of 2800 K. The mean electron and \HH\ fractions are now
0.03 and $4\times10^{-5}$: the latter is 20 times larger than the
primordial value, and will delay the buildup of a uniform UV
photodissociating background. The net effect of the X-rays is to
reduce gas clumping in the IGM by as much as a factor of 3. While the
suppression of baryonic infall and the photoevaporation of some halo
gas lower the gas mass fraction at overdensities $\delta$ in the range
20-2000, enhanced molecular cooling increases the amount of dense
material at $\delta>2000$. In many haloes within the proximity of our
miniquasar the \HH-boosting effect of X-rays is too weak to overcome
heating, and the cold and dense gas mass actually decreases.  We find
little evidence for an entropy floor in gas at intermediate densities
preventing gas contraction and \HH\ formation: we show, instead, that
molecular cooling can affect the dynamics of baryonic material before
it has fallen into the potential well of dark matter haloes and
virialized.  Overall, the radiative feedback from X-rays enhances gas
cooling in lower-$\sigma$ peaks that are far away from the initial
site of star formation, thus decreasing the clustering bias of the
early pregalactic population, but does not appear to dramatically
reverse or promote the collapse of pregalactic clouds as a whole.
\end{abstract}


\section{Introduction}

The first `seed' black holes (BHs) that later grew to become the
supermassive variety that power active galactic nuclei (AGNs) must
have appeared at very early epochs, $z>10$: this is in order to have
had sufficient time to build up via gas accretion a mass of several
$\times 10^9\,\Msun$ by $z=6.4$, the redshift of the most distant
quasars discovered in the {\it Sloan Digital Sky Survey} (Fan \etal
2003).  The origin and nature of this seed population remain
uncertain.  Numerical simulations performed in the context of
hierarchical structure formation theories show that the first stars
(the so-called `Population III') in the Universe formed out of
metal-free gas in dark matter `minihaloes' of mass above a few $\times
10^5\,h^{-1}\,\Msun$ (Abel, Bryan, \& Norman 2000; Fuller \& Couchman
2000; Yoshida \etal 2003; Reed \etal 2005) condensing from the rare
high-$\sigma$ peaks of the primordial density fluctuation field at
$z>20$, and were likely very massive (e.g. Abel, Bryan, \& Norman
2002; Bromm, Coppi, \& Larson 2002; see Bromm \& Larson 2004 for a
recent review).  Non-rotating very massive stars in the mass window
$150\lesssim m_*\lesssim 250\,\Msun$ are expected to disappear as
pair-instability supernovae (Bond, Arnett, \& Carr 1984) and leave no
compact remnants. Stars with $40\lesssim m_*\lesssim 150\,\Msun$ and
$m_*\gtrsim 250\,\Msun$ are predicted instead to collapse to BHs with
masses comparable to those of their progenitors (Fryer, Woosley, \&
Heger 2001). Barring any fine tuning of the initial mass function of
Population III stars, intermediate-mass BHs -- with masses above the
5--20$\,\Msun$ range of known `stellar-mass' holes -- may then be the
inevitable endproduct of the first episodes of pregalactic star
formation (Madau \& Rees 2001).  Another route for the creation of
more massive black hole seeds may be the formation -- and subsequent
collapse as a result of the post-Newtonian instability -- of
supermassive stars with $m_*\gg 10^3\,\Msun$ (see, e.g., Shapiro 2004a
for a recent review) out of the lowest angular momentum gas in rare
haloes above the minimum mass for atomic cooling at high redshifts
(Bromm \& Loeb 2003; Koushiappas, Bullock, \& Dekel 2004). Our
simulations are related more closely to the former type of scenario.

Physical conditions in the central potential wells of gas-rich
protogalaxies may have been propitious for BH accretion.  In the
absence of an \HH\ photodissociating UV flux and of ionizing X-ray
radiation, three-dimensional simulations of early structure formation
show that the fraction of cold, dense gas available for accretion on to
seed holes or star formation exceeds 20 per cent for haloes more massive than
$10^6\,\Msun$ (Machacek, Bryan, \& Abel 2003). There is also the
possibility that the collapse of a rotating very massive star may lead
to the formation of a massive BH and a `ready-made' accretion disc
(Fryer \etal 2001; Shapiro 2004b) that may provide a convenient source
of fuel to power a `miniquasar'.

The presence of accreting BHs powering Eddington-limited miniquasars
at such a crucial formative stage in the evolution of the Universe
presents a challenge to models of the epoch of first light and of the
thermal and ionization early history of the intergalactic medium
(IGM), and serves as the main motivation of this paper.  Feedback
processes from the first stars and their remnants likely played a key
role in reheating and structuring the IGM and in regulating gas
cooling and star formation in pregalactic objects. Energetic photons
from miniquasars may make the low-density IGM warm and weakly ionized
prior to the epoch of reionization breakthrough (Madau \etal 2004;
Ricotti, Ostriker, \& Gnedin 2005). X-ray radiation may boost the
free-electron fraction and catalyze the formation of \HH\ molecules in
dense regions, counteracting their destruction by UV Lyman-Werner
radiation (Haiman, Abel, \& Rees 2000; Machacek \etal 2003).  Or it
may furnish an entropy floor to the entire IGM, preventing gas
contraction and therefore impeding rather than enhancing \HH\
formation (Oh \& Haiman 2003).  Photoionization heating may evaporate
back into the IGM some of the gas already incorporated into haloes
with virial temperatures below a few thousand kelvins (Barkana \& Loeb
1999; Haiman, Abel, \& Madau 2001; Shapiro, Iliev, \& Raga 2004). The
detailed consequences of all these effects is poorly understood.
   
In this paper, we describe the results of fully 3D Eulerian
cosmological hydrodynamical simulations of the effect of the first
miniquasars on the thermal properties of the high-redshift IGM. The
focus of our investigation is not the modeling of the processes that
lead to gas accretion on to BHs, but rather the radiative feedback of
this population on their environment and on structure formation as a
whole. Since their host haloes form from the collapse of rare density
fluctuations, miniquasars powered by accreting BHs are expected to be
strongly clustered and highly biased tracers of the underlying dark
matter (DM) distribution. We use the adaptive mesh refinement (AMR)
technique to home in, with progressively finer resolution, on the
densest parts of the `cosmic web'. Rather than approximating
radiation fields as isotropic, we study the impact of a point-source
of X-ray radiation that starts shining in a rare high-$\sigma$ peak at
$z=21$, before a universal ionizing background is actually
established. The outline of this paper is as follows. In \S~2 we
describe our suite of numerical simulations. \S~3 presents general
results of the simulations, with and without X-rays, and on the
influence on the thermal and ionization state of the IGM of a
non-uniform radiation field in the optically thin limit. We discuss
the distance-dependent radiative feedback effect of a miniquasar on
the amount of primordial gas that can cool and collapse, thus becoming
available for star formation, in \S~4. Finally, we present our
conclusions in \S5.

\section{Simulations}

High-resolution hydrodynamics simulations of early structure formation
in $\Lambda$CDM cosmologies are a powerful tool to track in detail the
thermal and ionization history of a clumpy IGM and guide studies of
early reheating. We have used a modified version of \textsc{enzo}, an
adaptive mesh refinement (AMR), grid-based hybrid (hydro$+$N-body)
code developed by Bryan \& Norman (see http://cosmos.ucsd.edu/enzo/)
to solve the cosmological hydrodynamics equations, study the cooling
and collapse of primordial gas in DM haloes, and simulate the
large-scale effect of a miniquasar turning on at very high
redshift. All the results shown below assume a $\Lambda$CDM world
model with parameters $\Omega_M=0.3$, $\Omega_\Lambda=0.7$, $h=0.7$,
$\Omega_b=0.05$, $\sigma_8=0.85$, and $n=1$.

The primordial distribution of gas and DM is initialized in a comoving
box of 1 Mpc on a side using the linear power spectrum of Eisenstein
and Hu (1999). We first identify in a pure N-body simulation the
Lagrangian volume of the most massive DM halo at $z=25$. At this time
its total mass is $7\times 10^5\,\Msun$, corresponding to a 3.5 sigma
peak in the density fluctuation field. Because its mass is comparable
to the mass threshold for gas cloud formation by molecular cooling,
this halo will likely harbor the first collapsed baryonic object in
our simulated volume. It is this halo that we flag as the host of the
first massive BH. We then generate new initial conditions centered
around this density peak, consisting of two nested $128^3$ static
grids, of which the inner one covers the central 0.5 Mpc volume. The
DM density field is also sampled with $128^3$ particles in the inner
region, leading to a mass resolution of $m_{\rm DM}=2000\,\Msun$. This
ensures that haloes above the cosmological Jeans mass are well
resolved at all redshifts $z<21$.  At $z=15$, the five most massive
haloes in the box have between $5\times 10^3$ and $10^4$ DM particles.

\begin{figure}
\includegraphics[width=3.in]{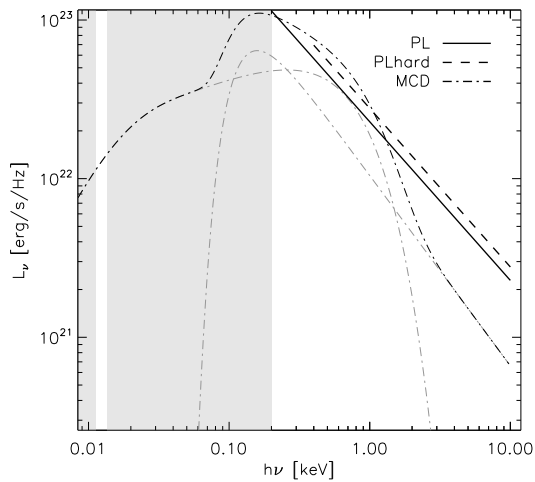}
\caption{
The three types of miniquasar spectra used in this paper. `PL'
(\textit{solid}) and `PLhard' (\textit{dashed}) are simple power
laws with $\alpha=1$ and lower-energy cutoffs of 0.2 and 0.4 keV,
respectively. `MCD' (\textit{dot-dashed}) is a combination of
multicolour disc and $\alpha=1.2$ power law. The two components are
shown separately in light grey. In the MCD case only we have included
radiation in the Lyman-Werner band ($11.2-13.6$ eV): photons between
13.6 eV and 0.2 keV are assumed to be absorbed within the emitting
host halo. All spectra have been normalized to the Eddington
luminosity of a $150\,\Msun$ black hole with radiative efficiency
$\epsilon=0.1$.
}
\label{fig:miniqso_spectra}
\vspace{+0.3cm}
\end{figure}

\begin{table*}
\begin{center}
\caption{
Summary of simulations.
}
\begin{tabular}{lccccl}
Name & Spectrum & Energy range & $z_{\rm on}$ & Lifetime & Description \\
\hline
NoBH & -- & -- & -- & -- & Reference simulation without miniquasar \\
PL & power law & 0.2-10 keV & 21-17.5 & 45 Myr & Fiducial simulation with miniquasar\\
PLlong & power law & 0.2-10 keV & 21-15.5 & 90 Myr & Extended miniquasar lifetime \\
PLhard & power law & 0.4-10 keV & 21-17.5 & 45 Myr & Harder spectrum \\
MCD & multicolour disc $+$ power law & -- & 21-17.5 & 45 Myr & Includes Lyman-Werner 
\HH-dissociating flux\\
\end{tabular}
\label{tab:summary}
\end{center}
\end{table*}

During the evolution from $z=99$ to $z=15$, refined (child) grids are
introduced with twice the spatial resolution of the coarser (parent)
grid. The AMR is restricted to the inner 0.5 Mpc and is triggered when
a cell reaches a DM overdensity (baryonic overdensity) of 2.0
(4.0). This low overdensity refinement criterion has been shown to
yield comparable results in equivalent simulations with \textsc{enzo} and the
smoothed particle hydrodynamics (SPH) code GADGET over the entire
range of the dark halo mass function (O'Shea \etal 2005a). The region
of interest is allowed to dynamically refine further to a total of 8
levels on a $128^3$ top grid, resulting in a maximum dynamic range
(ratio of box size to the smallest spatial scale that can be resolved)
of 32,768 or a maximum resolution of 30 pc (comoving). At the end of a
typical simulation, the code uses $\gtrsim 2.2\times 10^7$ computational
grid cells on $\gtrsim 9800$ grids, with a median number of cells per
grid of 768. 

The spatial distribution of $128^3\times 15/8$ collisionless dark
matter particles determines at every time-step the large-scale
gravitational field in the box. We evolve the non-equilibrium rate
equations for nine species (H, H$^+$, H$^-$, e, He, He$^+$, He$^{++}$,
\HH, and \HH$^+$) in primordial self-gravitating gas, including
radiative losses from atomic and molecular line cooling, Compton
heating by X-rays and Compton cooling by the cosmic background
radiation. The rate coefficients for the reaction network for these
species are primarily those described in Abel \etal (1997): the \HH\
cooling function is that of Lepp \& Shull (1983). We impose
the constraint that the minimum temperature attainable by radiative
cooling is that of the cosmic microwave background.

A zero-metallicity progenitor star of mass $m_*=150\,\Msun$ turning on
at $z=25$ in the host halo will emit $\gtrsim 10^{64}$ photons above 13.6
eV over a lifetime of $2.3\,$ Myr (Schaerer 2002), about 10 such
photons per halo baryon. The ensuing ionization front will overrun the
host halo, photoevaporating most of the surrounding gas (Whalen, Abel,
\& Norman 2004). Gas accretion on to the BH remnants of the first stars
that create \HII regions may have to wait for new cold material to be
made available through the hierarchical merging of many gaseous
subunits. The simulations show that, by $z=21$, our massive BH has
been incorporated into a larger DM halo that contains 7 times more gas
than the original host: it is at this time that we turn on the
miniquasar radiation field. Our miniquasar is powered by a $m_{\rm
BH}=150\,\Msun$ hole accreting at the Eddington rate and shining for 1
(2 in one of our 7 simulations) Salpeter time-scale,
$t_S=450\,\epsilon/(1-\epsilon)$ Myr, with a radiative efficiency of
$\epsilon=10$ per cent. Its exponentially growing mass and luminosity are
recomputed at every time-step, and its trajectory is followed by
flagging the DM particle corresponding to the maximum density in the
host halo as our BH. The location of this flagged particle is
determined at every time-step and used as the origin of an isotropic
$1/r^2$ radiation field: from $z=21$ to $z=15$ our BH typically moves
by less than 6 kpc (comoving). Note that, because of the considerable
mass difference between our DM particle resolution ($2000\,\Msun$) and
the initial mass of the black hole ($150\,\Msun$), we are not able to
accurately track the true trajectory of the hole.

The photon energy distribution of putative high-z miniquasars is
uncertain.  The spectra of `ultraluminous' X-ray sources in nearby
galaxies appear to require both a soft component (well fit by a cool
multicolour disc blackbody with $kT_{\rm max} \simeq 0.15$ keV, which
may indicated intermediate-mass BHs; Miller \& Colbert 2004) and a
non-thermal power-law component, $L_E \propto E^{-\alpha}$, of
comparable luminosity and slope $\alpha \sim 1$. Here we have run
simulations with three different spectra: one (`PL') assuming a
simple power-law miniquasar spectrum with $\alpha=1$ for photons with
energies in the range $0.2-10$ keV; one (`PLhard') with a similar
spectrum, but with a low-energy cutoff of 0.4 keV; and one (`MCD')
with luminosities equally divided between a multicolour disc component
and a power law with $\alpha=1.2$, both with a 0.2 keV low-energy
cutoff. In the multicolour disc component, each annulus of a thin
accretion disc is assumed to radiate as a blackbody with a radius
dependent temperature, $T(r)\propto r^{-3/4}$, and the temperature of
the innermost portion of the disc is related to the mass of the BH as
$T_{\rm in}\propto m_{\rm BH}^{-1/4}$ (e.g. Makishima \etal 2000). All
spectral energy distributions are normalized to the Eddington
luminosity, which is exponentially increasing during the active phase
of the miniquasar. A plot of the different input spectra is shown in
Fig.~\ref{fig:miniqso_spectra}.  For the MCD case only, we included
in addition to the X-rays an \HH-dissociating flux in the Lyman-Werner
band (LW, $11.2-13.6$ eV), with an intensity determined by the
Raleigh-Jeans tail of the multicolour disc. The implicit assumption for
all of these spectra is that photons with energies between 13.6 eV and
0.2 keV are absorbed within the miniquasar host halo. For this reason
our simulation is not strictly valid within the host halo, and we have
excised it from all of the following analysis.

\begin{figure*}
\includegraphics[width=6.in]{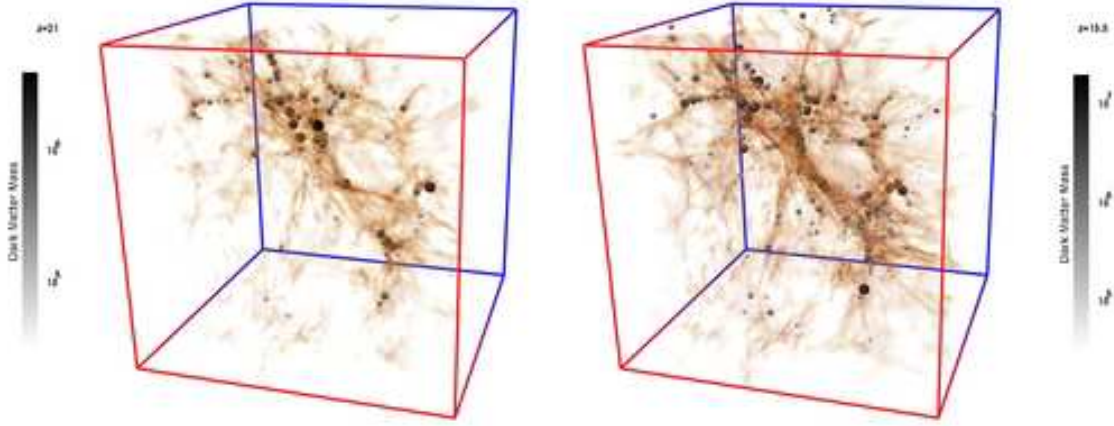}
\caption{
A 3D volume rendering of the IGM in the inner 0.5 Mpc simulated box at
$z=21$ and $z=15.5$. Only gas with overdensity $4<\delta<10$ is shown:
the locations of dark matter minihaloes are marked by spheres. The
size and grey scale of the spheres indicate halo mass. At these
epochs, the halo finder algorithm identifies 55 ($z=21$) and 262
($z=15.5$) minihaloes within the simulated volume.
}
\label{fig:3Dvol}
\end{figure*}

In the pre-reionization Universe, when the IGM is predominantly
neutral, soft X-ray photons will be absorbed as they photoionize
hydrogen or helium atoms. For a mixture of H and He with cosmic
abundances, the effective bound-free absorption cross-section can be
approximated to an accuracy of 30 per cent in the range $50\,{\rm
eV}<h\nu<2\,$keV as $\sigma_{\rm bf}\approx 4\times 10^{-20}\, {\rm
cm^2}~(h\nu/0.1\,{\rm keV})^{-3}$. The mean free path of ionizing
radiation in the neutral medium with overdensity $\delta\equiv
\rho/\rho_b$ is then
\be
\lambda={1\over n_\H\,\sigma_{\rm bf}}\approx 5\,{\rm kpc}\,
\left({1+z\over 20}\right)^{-3}\,\left({h\nu\over 0.1\,{\rm keV}}\right)^3\,\delta^{-1}.
\ee
Throughout the simulation box, gas at the mean density will be
transparent to photons with energies $>0.2$ keV and we take advantage
of this by working in the optically thin approximation. At $z=21$ even
the most massive haloes in our simulation have total hydrogen columns
below $10^{22}$ cm$^{-2}$, i.e. are transparent to photons above 0.7
keV. Since a power-law spectrum with $\nu\,I_\nu\sim$ const is
characterized by equal power per logarithmic frequency interval,
photoelectric absorption by intervening haloes will not significantly
attenuate the ionizing energy flux: this is then only a function of
distance from the miniquasar, and grows linearly with the mass of the
hole in the PL and PLhard cases.  Photo-ionizations and
photodissociations couple the radiation field to the hydrodynamics
through a heating term in the energy conservation equation and source
and sink terms in the species abundance equations (Anninos \etal
1997). The radiative heating $H_j$ and photoionization and
photodissociation $\Gamma_j$ rate coefficients are given by
\ba
H_j & = & \int_{\nu_{0,j}}^\infty \sigma_j(\nu) I(\nu)
{h\nu-h\nu_{0,j}\over h\nu}\,d\nu \\
\Gamma_j & = & \int_{\nu_{0,j}}^\infty \sigma_j(\nu)
{I(\nu)\over h\nu}\,d\nu,
\ea
where $\sigma_j(\nu)$ is the cross-section and $\nu_{0,j}$ is the
frequency threshold for the $j^{th}$ reaction, and $I(\nu)$ is the
intensity of the radial radiation field.

\begin{figure*}
\includegraphics[width=7.in]{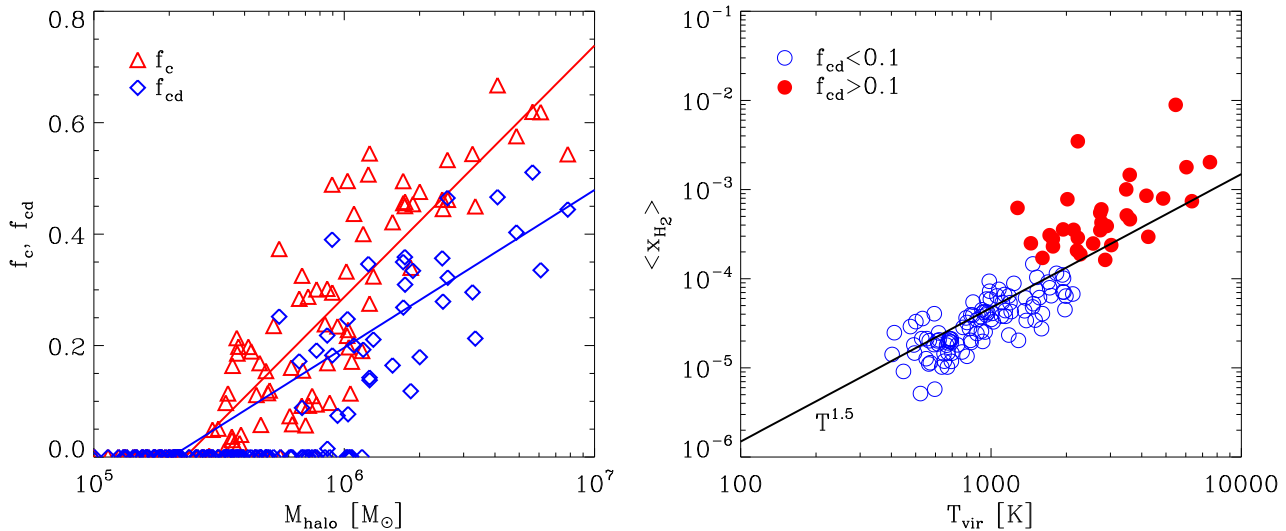}
\caption{
{\it Left:} Fraction of cold and cold$+$dense gas within the virial
radius of all haloes indentified at $z=17.5$ with $\Tvir>400$, as a
function of halo mass. {\it Triangles:} $f_c$, fraction of halo gas
with $T<0.5\,\Tvir$ and $\delta>1000$ that cooled via roto-vibrational
transitions of \HH.  {\it Diamonds:} $f_{\rm cd}$, fraction of gas
with $T<0.5\,\Tvir$ and $\rho> 10^{19}\,\Msun$ Mpc$^{-3}$ that is
available for star formation. The straight lines represent mean
regression analyses of $f_c$ and $f_{\rm cd}$ with the logarithm of
halo mass. {\it Right:} Mass-weighted mean \HH\ fraction as a function
of virial temperature for all haloes at $z=17.5$ with $\Tvir>400\,$K
and $f_{\rm cd}<0.1$ ({\it empty circles}) or $f_{\rm cd}>0.1$ ({\it
filled circles}). The straight line marks the scaling of the
temperature-dependent asymptotic molecular fraction.
}
\label{fig:coldense_noBH}
\end{figure*}

\section{General results}
\subsection{`NoBH' simulation}

A reference simulation with no radiation source (`NoBH') was run for
comparison.  The clustered structure around the selected high-$\sigma$
peak of the density field is clearly seen in Fig.~\ref{fig:3Dvol}, a
3D volume rendering of the inner 0.5 Mpc box at redshifts 21 and
15.5. The figure shows gas at $4<\delta<10$, with the locations of
dark matter minihaloes marked by spheres colored and sized according
to their mass (the spheres are only markers, the actual shape of the
haloes is typically non-spherical). Several interleaving filaments are
visible, at the intersections of which minihaloes are typically
found. We identify the locations of DM minihaloes in the simulation
volume by employing the HOP halo finder developed by Eisenstein \& Hut
(1998). This algorithm identifies haloes by grouping together
particles that are associated with the same local density maximum,
without enforcing spherical symmetry around the densest particle. We
consider only haloes containing more than 100 particles. At $z=21$, 55
haloes in our volume satisfy this criterion, by $z=17.5$ this number
has grown to 149, and by $z=15.5$ to 262 haloes. At this epoch, only
four haloes have reached the critical virial temperature for atomic
cooling, $\Tvir=10^4\,$K, where
$$
\Tvir=1.7\times 10^4 \left({\mu\over 1.2}\right) \left({M_{\rm halo}\over
10^7\,h^{-1}\,\Msun}\right)^{2/3} 
$$
\be
~~~~~~~~~~~~~~~~~~~\times \left[{\Omega_M\over \Omega_M(z)}\,{\Delta_{\rm 
vir}\over 18\pi^2}\right]^{1/3}\,\left({1+z\over 20}\right)\,{\rm K},
\ee 
$\mu$ is the mean molecular weight, and $\Delta_{\rm vir}$ is the
density contrast at virialization.

The primordial fractional abundance of \HH\ in the IGM is small,
$x_{\rm H_2}\approx 2\times 10^{-6}$, as at $z>100$ \HH\ formation is
inhibited because the required intermediaries, either \HH$^+$ or
H$^-$, are destroyed by cosmic microwave background (CMB) photons
(e.g. Galli \& Palla 1998). Most of the gas in the simulation
therefore cools by adiabatic expansion. Within collapsing minihaloes,
however, gas densities and temperatures are large enough that \HH\
formation is catalyzed by H$^-$ ions through the associative
detachment reaction H+H$^-\rightarrow$ \HH+$e^-$, and the molecular
fraction increases at the rate $dx_{\rm H_2}/dt\propto x_e n_{\rm HI}
\Tvir^{0.88}$, where $x_e$ is the number of electrons per hydrogen
atom. For $\Tvir \lesssim$ a few thousand kelvins the virialization shock
is not ionizing, the free electrons left over from recombination are
depleted in the denser regions, and the production of \HH\ stalls at a
temperature-dependent asymptotic molecular fraction $x_{\rm H_2}
\approx 10^{-8}\, \Tvir^{1.5}\,\ln(1+t/t_{\rm rec})$, where $t_{\rm
rec}$ is the hydrogen recombination time-scale (Tegmark \etal 1997). A
typical \HH\ fraction in excess of 200 times the primordial value is
therefore produced after the collapse of structures with virial
temperatures of order $10^3\,$K. This is large enough to efficiently
cool the gas and allow it to collapse within a Hubble time unless
significant heating occurs during this phase (Abel \etal 2000; Yoshida
\etal 2003).

Fig.~\ref{fig:coldense_noBH} (left panel) shows the fraction of cold
gas within the virial radius as a function of halo mass for all the
haloes identified at redshift 17.5. Note that the presence of halo
masses below the HOP selection limit of $2 \times 10^5\,\Msun$ is
caused by differences between the extent of a halo as defined by the
HOP algorithm and the virial radius measured by the
spherically-averaged radial profile analysis. Following Machacek \etal
(2001), we define $f_c$ as the fraction of gas with temperature
$<0.5\,\Tvir$ and density $>1000$ times the background (this is the
halo gas that is able to cool below the virial temperature because of
\HH), and $f_{\rm cd}$ as the fraction of gas with temperature
$<0.5\,\Tvir$ and (physical) density $>10^{19}\, \Msun$ Mpc$^{-3}$
(this is the self-gravitating gas available for star formation). As in
Machacek \etal (2001), we find that both $f_c$ and $f_{\rm cd}$ are
correlated with the halo mass, with the fraction of cold$+$dense gas
increasing less rapidly than $f_c$ with $M_{\rm halo}$. The mass
threshold for significant baryonic condensation (non-zero $f_{\rm
cd}$) is approximately $5\times 10^5\,\Msun$ at these redshifts
(Haiman, Thoul, \& Loeb 1996).  Also depicted in
Fig.~\ref{fig:coldense_noBH} (right panel) is the mass-weighted mean
molecular fraction of all haloes with $\Tvir>400\,$K. Filled circles
represent haloes with $f_{\rm cd}>0.1$, while open circles represent
the others. The straight line marks the scaling of the asymptotic
molecular fraction in the electron-depletion transition regime. Haloes
with significant cold$+$dense gas fraction have larger $\langle x_{\rm
H_2}\rangle$ than those shown in fig.~3 of Yoshida \etal
(2003). This is due to the high resolution attainable by an AMR code
compared to SPH. The maximum gas density reached at redshift 15 in the
most refined region of our simulation is $4\times 10^5\,$cm$^{-3}$
(corresponding to an overdensity of $3\times 10^8$): within this cold
pocket the excited states of \HH\ are in LTE and the cooling time is
nearly independent of density (e.g. Lepp \& Shull 1983).

\begin{figure*}
\includegraphics[width=5.in]{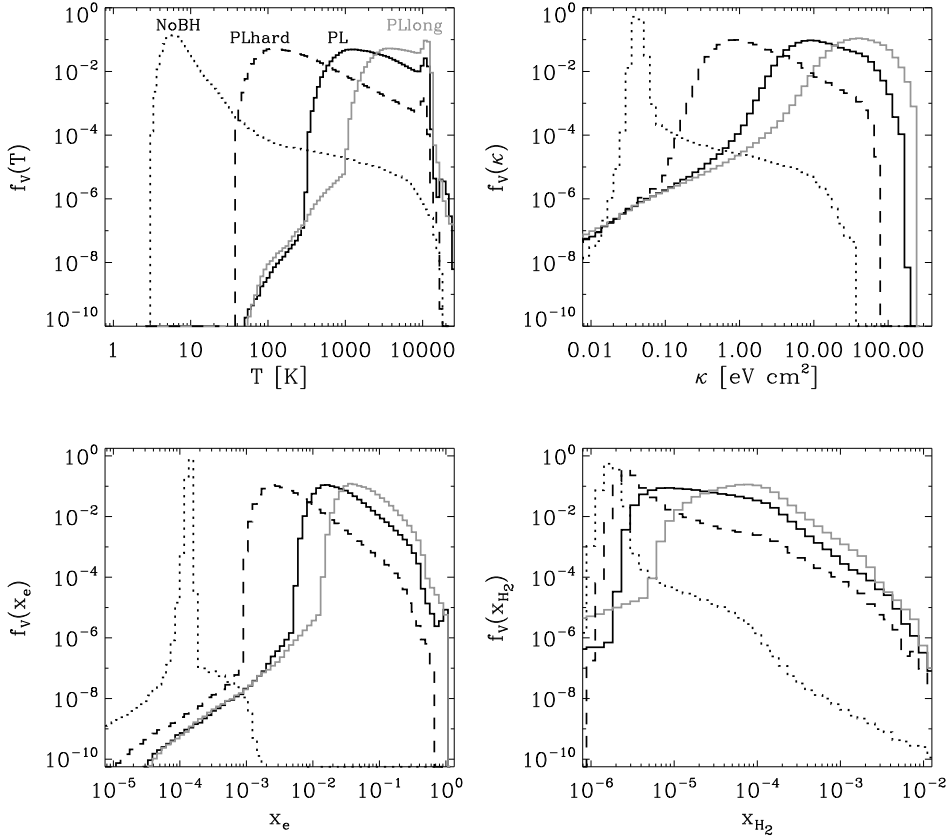}
\caption{
Distributions of temperature $T$ (binsize $\Delta\!\log T=0.04$),
`entropy' $\kappa$ ($\Delta\!\log \kappa=0.08$), electron fraction
$x_e$ ($\Delta\!\log x_e=0.07$), and molecular hydrogen fraction
$x_{\rm H_2}$ ($\Delta\!\log x_{\rm H_2}=0.10$) (clockwise from upper
left). \textit{Dotted lines:} $z=17.5$ NoBH run without miniquasar
feedback.  \textit{Dark solid lines:} $z=17.5$ fiducial simulation, in
which the miniquasar shines for one Salpeter time
($z=21-17.5$). \textit{Light solid lines:} $z=15.5$ PLlong run, in
which the miniquasar shines for two Salpeter times ($z=21-15.5$).
\textit{Dashed lines:} $z=17.5$ PLhard run, in which the low-energy
cutoff of the miniquasar spectrum has been increased from 0.2 to 0.4
keV. The volume-averaged temperature, entropy, electron fraction, and
molecular fraction are $(8\,{\rm K}, 0.04\,{\rm eV~cm^2}, 1.4\times
10^{-4}, 1.9\times 10^{-6})$, $(2800\,{\rm K}, 17\,{\rm eV~cm^2},
0.03, 3.9\times 10^{-5})$, $(6150\,{\rm K}, 49\,{\rm eV~cm^2}, 0.07,
9.3\times 10^{-5})$, and $(383\,{\rm K}, 2.2\,{\rm eV~cm^2}, 0.005,
6.3\times 10^{-6})$, respectively.
}
\label{fig:distributions}
\end{figure*}

At $z=17.5$, $\sim 90$ per cent of the simulated volume contains gas with
$T<10\,$ K, as shown by the differential temperature distributions in
Fig.~\ref{fig:distributions}. The distribution of the entropy
parameter $\kappa=kTn^{-2/3}$ is even more strongly peaked than the
temperature distribution, as entropy is conserved during infall on to
non-linear structures or Hubble expansion, which heat/cool the gas
adiabatically. After decoupling from the CMB, the IGM temperatures
drops as $T(z)\simeq 2.73 (1+z_d) [(1+z)/(1+z_d)]^2\,$K for
$z<z_d=150$, giving rise to an entropy floor of $0.04$
eV cm$^2$, independent of redshift. Only gas that is shock heated
during accretion on to DM haloes or that cools radiatively via \HH\ can
depart from this peak. Shock heating produces the extended tail out to
$\kappa=10$ eV cm$^2$ at a volume fraction in the range
$f_V=10^{-5}-10^{-4}$. At these epochs, we find a collapsed gas mass
fraction of $f_M=2.5$ per cent. The fraction of baryons that gets shock
heated ($\kappa>0.1$ eV cm$^2$) is $f_M=6.7$ per cent. Radiative cooling
processes affects such a tiny fraction of the volume that it remains
outside the plotted range.\footnote{The effect of radiative cooling
can be observed in the `phase diagram' of
Fig.~\ref{fig:T_rho_phase} as the bluish tail at $\delta>100$,
$T<1000$ K, and volume fraction $f_V<10^{-8}$.}\ Due to the lack of
ionizing sources the medium is almost completely neutral, with a
distribution of electron fractions sharply peaked at $x_e=1.5\times
10^{-4}$. Note that the distribution of molecular hydrogen fraction,
$x_{\rm H_2}$, is considerably broader than that of $x_e$.

\begin{figure*}
\includegraphics[width=5in]{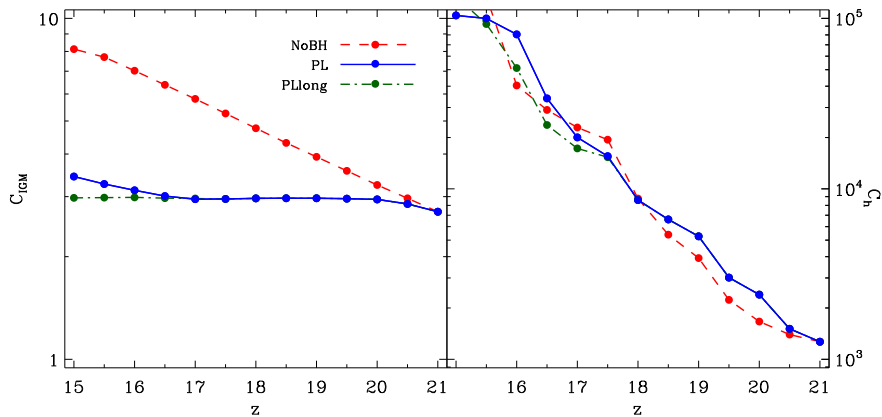}
\caption{
Clumping factor ($\langle n^2 \rangle/\langle n \rangle^2$) as a
function of redshift for the NoBH, PL, and PLlong simulations. {\it
Left:} Evolution of the clumping factor associated with diffuse
`intergalactic' ($\delta<70$) material. {\it Right:} halo
($\delta>70$) clumping factor. Note that this is not reduced in the
presence of X-ray heating.
}
\label{fig:clumpingfactor}
\end{figure*}

Another quantity of interest is the gas clumping factor $C=\langle n^2
\rangle/\langle n\rangle^2$, where the brackets denote volume-averaged
quantities.  The shorter radiative recombination time-scale, $t_{\rm
rec}=(\alpha \langle n_e\rangle C)^{-1}$ of a clumpy medium relative
to a homogeneous one increases the total number of photons per baryon
required for reionization. We have calculated the average clumping
factor within our simulation box of volume $L^3$ as
\be
C=C_{\rm IGM}+C_h=\sum_{\delta<70} {n_i^2 V_i\over \langle n_H\rangle^2 L^3}+
\sum_{\delta>70} {n_i^2 V_i\over \langle n_H\rangle^2 L^3},
\ee
where $n_i$ is the number density of hydrogen in the $i^{th}$ grid
cell of volume $V_i$, $\langle n_H\rangle$ is the background hydrogen
density, and the two sums are taken over all grid cells with baryonic
overdensity smaller and larger than $70$. We use this density
threshold -- corresponding to the mean baryonic overdensity at the
virial radius of our haloes -- to differentiate dense gas belonging to
virialized haloes ($C_h$) and diffuse uncollapsed `intergalactic'
material ($C_{\rm IGM}$). Gas within haloes only contributes to
recombinations when it is photoionized: self-shielding of a cold
gaseous disc and photoevaporation of halo gas by an external radiation
field all make the contribution of dense material with $\delta \gtrsim
70$ to the overall clumping very uncertain (e.g. Haiman \etal
2001). In Fig.~\ref{fig:clumpingfactor} we have plotted the
evolution with redshift of $C_{\rm IGM}$ and $C_h$ in the NoBH
simulation.  The clumping factor increases rapidly with time as
structure formation progresses, with $C_{\rm IGM}\approx 2.65\exp
[0.2(21-z)]$ over the range in redshift we probe.

\begin{figure*}
\includegraphics[width=6.5in]{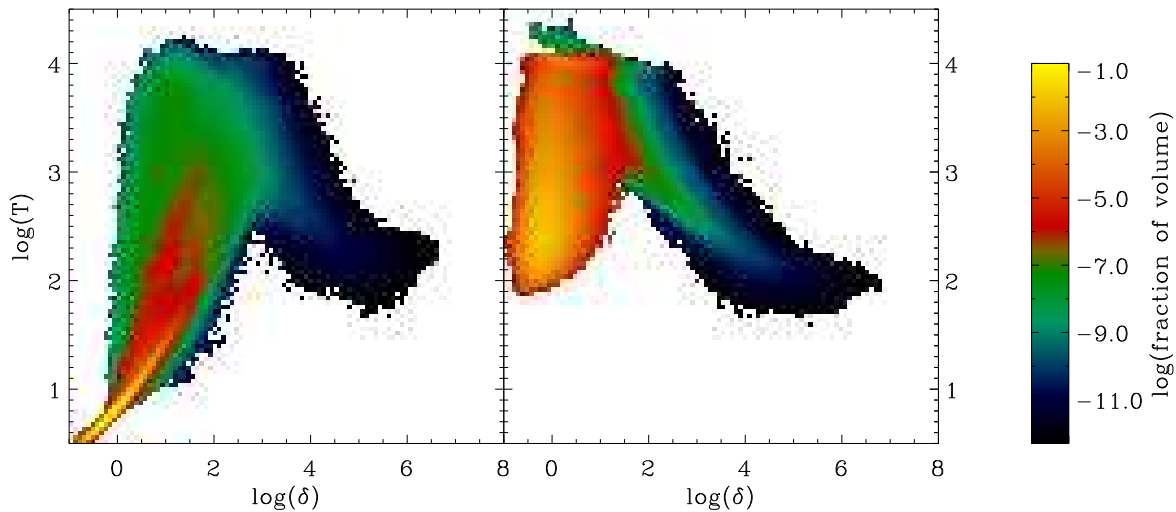}
\caption{
Two-dimensional distribution of gas temperature versus baryonic
overdensity at $z=17.5$. The color coding indicates the fraction of the
simulated volume at a given ($\delta,T)$. {\it Left:} NoBH run. 
{\it Right:} PL run.  
}
\label{fig:T_rho_phase}
\end{figure*}

\subsection{Heating and ionization by the miniquasar}
\label{sec:heating}

X-ray radiation from the miniquasar partially ionizes most of the gas
in the simulated volume both by direct photoionization of \HI and \HeI
and indirectly by collisional ionization of \HI by the fast
photolectrons. We have adopted the fitting formulae of Shull \& van
Steenberg (1985), cast as a function of the hydrogen ionized fraction,
$x=n_{\H^+}/n_{\H}$, to determine the fraction of energy of a
photoelectron deposited as heat versus further ionization. For photon
energies above 0.2 keV, the ratio of the \HI/\HeI photoionization
cross-sections drops below 4 per cent: since the primordial ratio of helium
to hydrogen is about 8 per cent, the photoionization and heating rates are
dominated by helium absorption, which exceeds the hydrogen
contribution by $\sim 2:1$. While \HeI is the main source of hot
primary photoelectrons, however, it is \HI that undergoes the bulk of
secondary ionizations. A primary nonthermal photoelectron of energy
$E=1\,$keV in a medium with residual ionization fraction (from the
recombination epoch) $x=2\times 10^{-4}$ will create over two dozens
secondary electrons, depositing a fraction $f_1=0.37$ of its initial
energy as secondary ionizations of \HI, $f_2=0.05$ as secondary
ionizations of \HeI, and $f_3=0.13$ as heat. The time-scale for
electron-electron encounters resulting in a fractional energy loss
$f=\Delta E/E$,
\be
t_{\rm ee}\approx 140\, {\rm yr}~Ef\,\left({1+z\over 20}\right)^{-3}\,
\left({\ln\Lambda\over 20}\right)^{-1}\,x^{-1}\,\delta^{-1}
\ee
(where $E$ is measured in keV), is typically much shorter than the
electron Compton cooling time-scale: the primary photoelectron will
therefore ionize and heat the surrounding medium before it is cooled
by the CMB.  The fraction of primary energy going into heat increases
gradually with $x$, reaching $f_3=0.5$ at $x=0.04$. Secondary
ionizations enhance the rates for ionization of \HI and \HeI (but not
for \HeII, Shull \& Van Steenberg 1985), and introduce an additional
coupling between the abundance rate equations. The rate coefficients
for \HI and \HeI ionizations can be written as
\be
\Gamma_{\rm HI}+f_1(x)\left(H_{\rm HI}+{n_{\rm HeI}\over n_{\rm HI}}\,H_{\rm HeI}\right)
{1\over 13.6\,{\rm eV}}
\ee
and
\be
\Gamma_{\rm HeI}+f_2(x)\left(H_{\rm HeI}+{n_{\rm HI}\over n_{\rm HeI}}\,H_{\rm HI}\right)
{1\over 24.6\,{\rm eV}},
\ee
respectively. We self-consistently evolve the hydrogen ionization
fraction $x$ and the hydrogen and helium abundances throughout the
simulations (note that the Shull \& van Steenberg fitting formulae
were derived assuming equal ionization fractions of hydrogen and
singly-ionized helium).

The two-dimensional distribution of gas overdensity and temperature at
$z=17.5$ is shown in Fig.~\ref{fig:T_rho_phase} for the NoBH and PL
simulations. The color coding in this phase diagram indicates the
fraction of the simulated volume at a given ($\delta,T)$. Most of the
gas in the NoBH run lies on the yellow line
($\kappa=KTn^{-2/3}T=const$) representing the initial adiabat. At low
overdensities the temperature either drops because of Hubble expansion
or rises because of adiabatic compression until the gas is shock
heated (red and green swath) to virial values, $T=10^3-10^4\,$K.  At
higher densities, the blue cooling branch follows the evolutionary
tracks in the temperature-density plane for spherically collapsing
clouds (Yoshida \etal 2003). \HH\ line emission lowers the temperature
down to $\approx 100\,$K, the minimum value attainable by
molecular cooling. The onset of the gravitational instability can
further compress the gas and cause a modest rise in temperature again
(Abel \etal 2000).

\begin{figure*}
\includegraphics[width=5.5in]{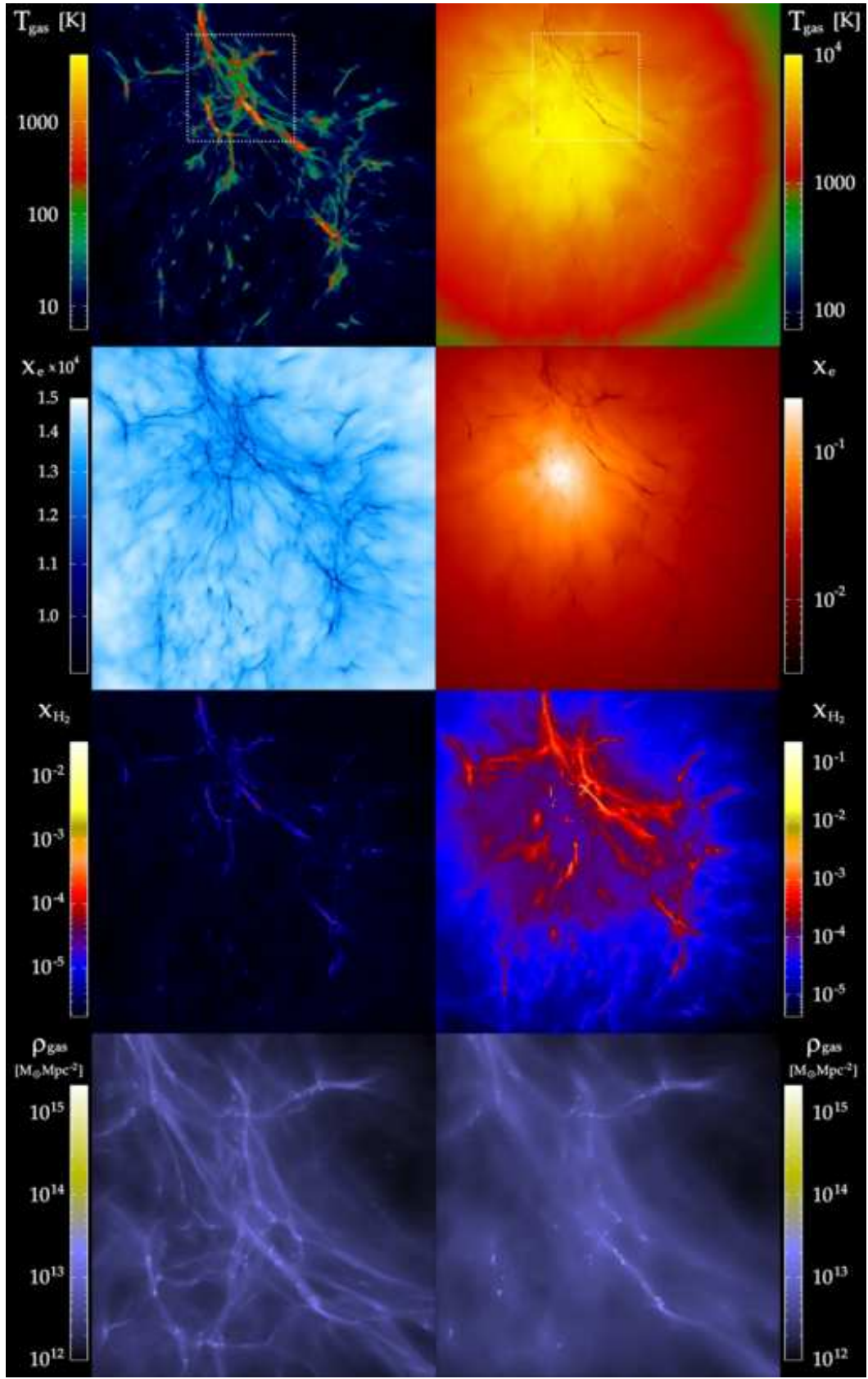}
\caption{{\it From top to bottom:} Projected temperature, electron
fraction, and molecular fraction distributions in the simulation box
for runs NoBH ({\it left}) and PL ({\it right}) at $z=17.5$. The
bottom panel shows the projected gas density in a smaller 127
comoving kpc region (dotted outline in top panels) to highlight the
effect of Jeans smoothing.}
\label{fig:Tcomp}
\end{figure*}

\begin{figure}
\includegraphics[width=3in]{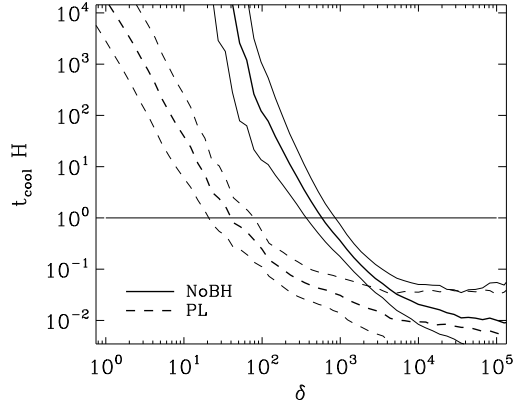}
\caption{
Mean \HH\ cooling time (in units of $H^{-1}$) versus gas overdensity,
for the NoBH ({\it solid line}) and PL ({\it dashed line}) runs at
$z=17.5$. The thin lines show departures of one root mean square
($\left(\langle t_{\rm cool}^2 \rangle - \langle t_{\rm cool}
\rangle^2\right)^{1/2}$) from the mean. Gas below the thin horizontal
line can cool down in a Hubble time.
}
\label{fig:tcool}
\end{figure}

After shining for a Salpeter time-scale, the miniquasar has heated up
the box to a volume-averaged temperature of 2800 K. The mean electron
fraction and entropy are now 0.03 and 17 eV cm$^2$ (see
Fig.~\ref{fig:distributions}): farther than 20 kpc (comoving) from the
source hydrogen is never ionized to more than 30 per cent. Gas near the
miniquasar is heated above $10^4\,$K and quickly cools down via
efficient atomic processes: the green finger at $\log \delta=0, \log
(T/{\rm K})>4$ represents baryonic material in this phase. The
increased electron fraction and gas temperature boost the gas-phase
\HH\ production, which occurs on a time-scale,
\be
t_{\rm H_2}= 30\,{\rm Myr}\,\left({x_{\rm H_2}\over 10^{-5}}\right) 
\left({0.01 \over x_e}\right) x_{\rm HI}^{-1} T_3^{-0.88}
\delta^{-1}\left({1+z\over 20}\right)^{-3}, 
\ee
that is much shorter than the hydrogen recombination time,
\be
t_{\rm rec}=9\times 10^3\,{\rm Myr}\,\left({0.01 \over x_e}\right) T_3^{0.64}
\delta^{-1}\left({1+z\over 20}\right)^{-3} 
\ee
(here $T_3\equiv T/10^3\,{\rm K}$).  The volume-averaged molecular
fraction is now 20 times larger than the primordial value, with denser
filaments in the IGM ($\delta\sim 10-20$) being traced out by a
smaller electron fraction and by a molecular fraction in the range
$10^{-5}-10^{-3}$ (see Fig.~\ref{fig:Tcomp}). This large molecular
fraction will hinder the buildup of a uniform UV photodissociating
background, as the maximum optical depth of the IGM in the \HH\ LW
bands can now exceed 10 (cf. Ricotti, Gnedin, \& Shull 2001; Haiman
\etal 2000). It will also promote \HH\ radiative cooling in filaments
on a time-scale,
\be
t_{\rm cool}\approx 650\,{\rm Myr}\,\left({10^{-3}\over x_{\rm H_2}}\right) 
x_{\rm HI}^{-1} \delta^{-1}\left({1+z\over 20}\right)^{-3}\,e^{730\,{\rm K}/T}, 
\ee
(Machacek \etal 2001) that can be shorter than the Hubble expansion
time, $1/H\simeq 285\,{\rm Myr}\,[(1+z)/20]^{-3/2}$, and the Compton
cooling time for mostly neutral primordial gas, $t_C=8\,{\rm
Myr}\,x_e^{-1}\,[(1+z)/20]^{-4}$. In Fig.~\ref{fig:tcool} we plot
the mean \HH\ cooling time in units of the expansion time-scale
$H^{-1}$ versus overdensity, at $z=17.5$. In the NoBH simulation only
gas with $\delta>600$ can cool in a Hubble time. This critical cooling
density is lowered to $\delta\approx 40$ in the PL simulation, due to
the enhanced \HH\ fraction. Molecular cooling can then affect the
dynamics of baryonic material before it has fallen into the potential
well of DM haloes and virialized. Gas in the simulation box is now
able to collapse and cool more rapidly within subsequent star-forming
minihaloes, without the need to produce much additional \HH.

\begin{figure}
\includegraphics[width=3in]{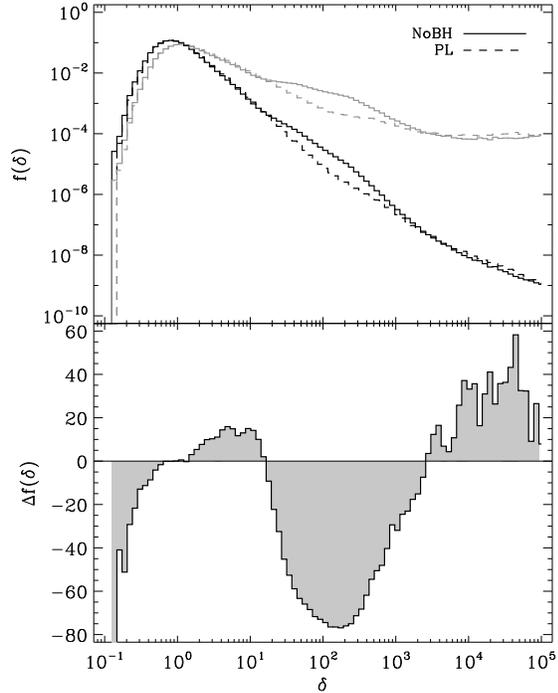}
\caption{
\textit{Top}: volume (\textit{black lines}) and mass (\textit{gray
lines}) fractions of gas overdensity $\delta=\rho/\rho_b$, in the NoBH
(\textit{solid lines}) and PL (\textit{dashed lines}) simulations at
$z=17.5$ (binsize $\Delta\!\log\delta=0.09$). \textit{Bottom}:
Percentage change, $100\times [f({\rm PL})-f({\rm NoBH})]/f({\rm
NoBH})$] in gas volume/mass fraction as a function of $\delta$.
}
\label{fig:densefraction}
\end{figure}

The global environmental impact of the miniquasar is illustrated in
Fig.~\ref{fig:densefraction}, where we plot in the top panel the
volume ($f_V$) and mass fraction ($f_M$) as a function of overdensity
in the NoBH and PL runs. The bottom panel shows the corresponding
percentage change from the NoBH to the PL case ($100 \times (f_{\rm
PL}-f_{\rm NoBH})/f_{\rm NoBH}$). Three prominent features are clearly
seen in this figure: 1) over most of the simulation volume, the net
effect of X-rays is to reduce gas clumping due to the smoothing of
sheets and filaments in the `cosmic web' by gas pressure. This
`Jeans smoothing' removes gas from intermediate $\delta$'s and puffs
it up to low overdensities ($\Delta f>0$ for $1<\delta<10$), thereby
filling in underdense regions ($\Delta f<0$ for
$\delta<1$);\footnote{In the miniquasar simulations the IGM clumping
factor remains approximately constant, $C_{\rm IGM}\simeq 3$, from
$z=21$ to $z=15$ (see the left panel in
Fig.~\ref{fig:clumpingfactor}). Because of the decreasing background
density with time, the IGM recombination time-scale at a given
temperature therefore increases as $(1+z)^{-3}$ over this redshift
interval.}\ 2) the suppression of baryonic infall and the
photoevaporation back into the IGM of some of the gas already
incorporated into haloes both lower the gas mass fraction at
$20<\delta<2000$; and 3) enhanced molecular cooling increases the
amount of dense material at $\delta>2000$. While the vast majority of
the baryons, both in volume and mass, are heated up, the densest gas
at the centres of haloes is actually cooled, due to the X-ray
catalysis of molecular hydrogen.  The feedback effect of X-rays on the
formation and cooling of pregalactic minihaloes will be discussed in
more details in the next section.

In the PLlong simulation we let the miniquasar shine for two Salpeter
times, from $z=21$ to $z=15.5$, leading to a final black hole mass of
$1100\,\Msun$. The increased intensity and duration of the radiation
further enhances the effects described above (see
Fig.~\ref{fig:distributions}). At $z=15.5$, the volume averages in the
box are: $\langle T \rangle=6150\;$K, $\langle \kappa \rangle=49\;{\rm
eV cm^2}$, $\langle x_e \rangle=0.07$, and $\langle x_{\rm H_2}
\rangle=9.3 \times 10^{-5}$. The additional heating, however, does not
further decrease the IGM clumping factor, and the halo clumping factor
is still not affected (see Fig.\ref{fig:clumpingfactor}). The increase
in $\langle x_{\rm H_2} \rangle$ by more than a factor of two is the most
significant difference between the PL to PLlong simulation. This will
make it even harder for subsequent generation of sources to establish
a uniform Lyman-Werner background. Note that if early black holes
undergo a period of super-Eddington mass growth (Volonteri \& Rees
2005; Haiman 2004), a possible subsequent luminous miniquasar phase,
powered by a more massive hole, might have feedback effects more
similar to, or even exceeding those of our PLlong simulation.

In order to probe the dependence of our results on the hardness
of the radiation field, we ran the PLhard simulation, in which the
lower energy cutoff was raised to $\nu_{\rm min}$=0.4 keV. Normalizing
the total energy output to $L_{\rm Edd}$ leads to a specific radiation
intensity which is 20 per cent higher (for $\alpha=1$) compared to the
PL simulation. The heating rate, on the other hand, scales roughly as
$\nu_{\rm min}^{-3}$. A factor of two increase in $\nu_{\rm min}$ thus
leads to a factor of eight decrease in the heating rate, and we expect
the feedback from the miniquasar to be much weaker. This expectation
is confirmed by the PLhard simulation, as shown by the dashed
histogram in Fig.~\ref{fig:distributions}. At $z=17.5$ the
volume-averaged temperature is only 380 K, with $\langle x_e \rangle =
0.005$ and $\langle x_{\rm H_2} \rangle = 6.2 \times 10^{-6}$. By the
same arguments we would expect the miniquasar feedback to strengthen
if the spectrum extended to lower energies, although the optically
thin assumption would then break down.

\section{X-ray radiative feedback: positive or negative?}
\label{sec:haloes}

The impact of the radiation produced by Population III stars and their
remnants on \HH\ chemistry and the properties of star-forming gas in
low-mass haloes has been studied by many authors. Molecular hydrogen
is fragile and easily photodissociated by soft UV photons below 13.6
eV (e.g. Haiman, Rees, \& Loeb 1997; Ciardi, Ferrara, \& Abel 2000;
Machacek \etal 2001; Glover \& Brand 2001).  While this negative
feedback may suppress gas cooling and collapse, the formation of \HH\
molecules is catalyzed by free electrons, and any process that
increased their abundance would boost the abundance of \HH\ as
well. Fossil \HII regions (Ricotti \etal 2001) and X-ray photons may
provide such positive feedback (Haiman \etal 2000; Venkatesan, Giroux,
\& Shull 2001; Machacek \etal 2003; Glover \& Brand 2003; Cen
2003). The issue of whether positive feedback processes dominate over
negative feedback is still open. All we offer here is an assessment of
the effect of X-rays in the absence of a strong photodissociating
flux. This condition would naturally be fulfilled if accreting
intermediate-mass BHs were the endproduct of the first episodes of
Population III star formation, as miniquasars are much more efficient
sources of radiation than their stellar-progenitors (Madau \etal
2004). As already mentioned in \S\,2, our MCD run includes in addition
to the X-rays a LW \HH-dissociating flux with an intensity determined
by the Raleigh-Jeans tail of the multicolour disc, $I_{\rm
LW}(z=21)=10^{-21}\,$ergs cm$^{-2}$ s$^{-1}$ Hz$^{-1}$ sr$^{-1}$ at a
comoving distance of 7.2 kpc, which is 10 (3) times lower than the 0.2
keV (1 keV) X-ray flux (cf. Machacek \etal 2003). We find the results
of the MCD simulation to be very similar to those of the PL run, and
it is the latter that we analyze in detail below.

\begin{figure*}
\includegraphics[width=6.0in]{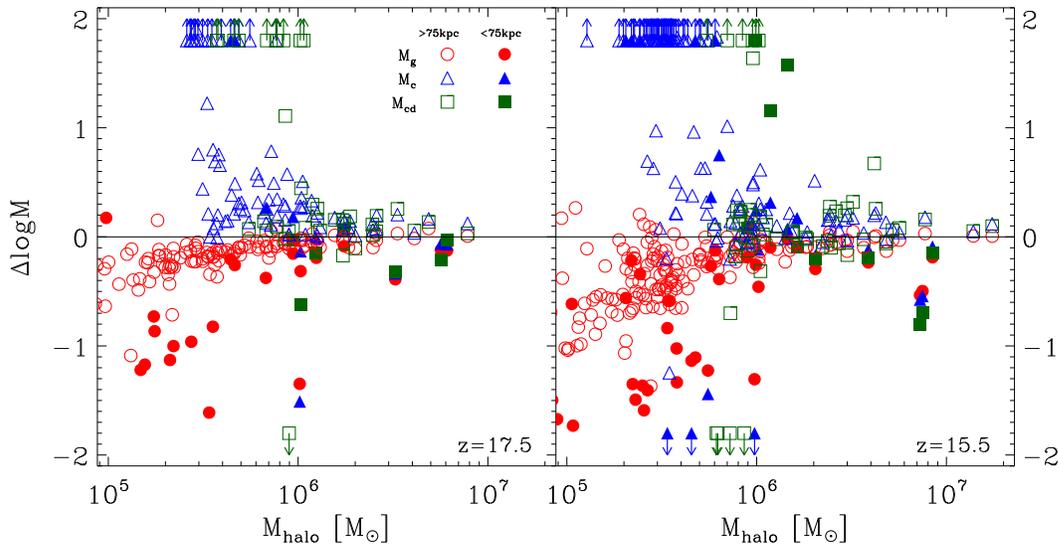}
\caption{
Variation in halo gas mass between the PL and NoBH simulations as a
function of halo total mass (as measured in the NoBH run) at two
different epochs.  Symbols show $\log [M_i({\rm PL})/M_i({\rm
NoBH})]$, with $i=g,c,cd$ for total gas, cold gas, and cold$+$dense
gas mass. Filled symbols represent haloes within 75 comoving kpc of
the miniquasars, empty symbols the others. Haloes with $|\Delta\log
M_{\rm c,cd}|>2$ are shown as arrows at the top and bottom of the
plot.
}
\label{fig:deltalogM}
\end{figure*}

Fig.~\ref{fig:deltalogM} depicts the change in halo gas content
between the PL and NoBH simulations as a function of halo mass at two
different epochs, $z=17.5$ (when the miniquasar stops shining) and
$z=15.5$: haloes in the two simulations have been matched by location
and DM mass. Different symbols represent total, cold, and cold$+$dense
gas mass both for haloes closer than 75 comoving kpc from the
radiation source and for those farther away. X-ray heating lowers the
total gas mass in nearly all virialized DM clumps, both by
photoevaporating baryons that were already incorporated into haloes
and by suppressing baryonic infall into newly forming structures. This
decrease in gas mass is more apparent below $M_{\rm halo}\approx
10^6\,\Msun$, and it is the strongest in pregalactic clouds close to
the miniquasars that have been exposed to the highest level of X-ray
flux.  Only a small fraction of the gas accreted by haloes in the NoBH
run is retained by the host gravitational potential in the PL run. One
of the potential wells that experiences the largest decrease in gas
content is fairly massive, $M_{\rm halo}=10^6\,\Msun$, is located only
12 kpc (comoving) away from the miniquasar, and has a total gas
fraction of 0.7 per cent, more than 20 times lower than in the NoBH
simulation!

\begin{figure}
\includegraphics[width=3.25in]{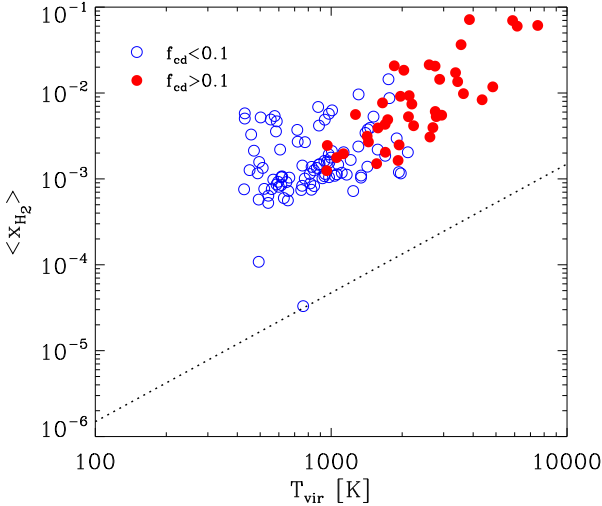}
\caption{
Same as Fig.~\ref{fig:coldense_noBH} ({\it right panel}), but for the
PL simulation. The dotted line is the $T^{1.5}$ line from the NoBH
case.
}
\label{fig:xH2_PL}
\end{figure}

Population III stars, however, can only form from gas that is able to
cool and condense. Fig.~\ref{fig:xH2_PL} shows the mass weighted
mean \HH\ fraction inside haloes in the PL simulation at $z=17.5$
(cf. right panel of Fig.~\ref{fig:coldense_noBH}). Overall $\langle
x_{\rm H_2} \rangle$ increases by slightly more than one order of
magnitude, independently of $\Tvir$. Compared with the NoBH case,
haloes with $f_{\rm cd}>0.1$ can be found at slightly lower
$\Tvir$. None of the haloes with $\Tvir<800 $K has any cold+dense
gas, even though their molecular fraction is quite high, $\langle
x_{\rm H_2} \rangle \approx 10^{-3}$. These haloes are the least massive
haloes in our simulation and were formed only after the miniquasar
began to shine. Continuous X-ray heating has prevented the gas from
cooling and falling into the halo. As a consequence their mean gas
mass fraction ($M_{\rm gas}/M_{\rm tot}$) is only 0.08 and their mean
central temperature exceeds their virial temperature by a factor of
3. Haloes with larger $\Tvir$, in contrast, formed earlier and their
gas has contracted to densities where \HH\ cooling is able to overcome
X-ray heating.

Evidence for X-ray enhanced cooling is seen in many haloes above
$2\times 10^5\,\Msun$. The largest relative increase in the amount of
cold material occurs in the mass range $2\times 10^5\,<M_{\rm
halo}<10^6\,\Msun$, where the boosting effect can exceed 1-2 orders of
magnitude! In this sense, X-ray preheating appears to somewhat
decrease the threshold mass required for efficient gas cooling and
star formation, albeit not by a large factor. In more massive peaks
most baryons are already cold in the absence of X-rays (see
Fig.~\ref{fig:coldense_noBH}), and positive feedback can only promote
a little additional cooling. Yet in many haloes in the proximity of our
miniquasar the \HH\-boosting effect of X-rays is too weak to overcome
heating, and the cold and dense gas mass actually decreases.  Table
\ref{tab:feedback} summarizes the miniquasar's feedback effect on the
total, cold, and cold+dense gas mass. We find evidence for a global
positive feedback in minihaloes more than 75 comoving kpc away from
the miniquasar, as the total mass in cold and cold$+$dense baryons
increase by 30 and 44 per cent, respectively. Farther than 150 kpc this
positive feedback is slightly larger yet, reaching $+ 50$ per cent for
cold gas. Within 75 kpc of the miniquasar, however, the feedback is
negative, as the total mass in cold$+$dense baryons decreases by more
than 50 per cent. The total amount of gas mass in haloes is reduced at all
distances within our simulated volume, with haloes closer than 75 kpc
losing on average more than half their gas mass. These effects are
long-lasting, and persist at $z=15.5$, 40 Myr after the miniquasar has
stopped shining (see right panel of
Fig.~\ref{fig:deltalogM}). Reducing the total amount of gas in the
outer regions of haloes, while raising the amount of cold and dense
gas available for star formation in the centres, could make it easier
for these haloes to chemically enrich their surroundings.

\begin{table}
\caption{
Summary of miniquasar feedback on halo total, cold, and cold+dense gas
mass, at $z=17.5$. All numbers are per cent change from NoBH to PL case.
}
\begin{tabular}{ccccc}
 & 0.5Mpc box & $<75$kpc & $>75$kpc & $>150$kpc \\
\hline
total & $-20.5$ & $-57.4$ & $-10.5$ & $-8.2$ \\
cold & $+19.4$ & $-43.4$ & $+43.5$ & $+49.3$ \\
cold+dense & $+5.7$ & $-52.1$ & $+29.3$ & $+38.4$ \\
\end{tabular}
\label{tab:feedback}
\end{table}

Overall, however, positive and negative feedbacks nearly compensate
each other, and the net impact of X-rays on the total amount of cold
material available for star formation within the simulated volume is
remarkably mild. By $z=17.5$, exposure to X-ray radiation has boosted
the total mass fractions of cold and cold$+$dense gas by only 14 and
6 per cent, respectively. There is hardly an X-ray `sterilizing effect':
star formation can proceed in minihaloes above the threshold mass,
including those that are near the radiation source. In the most
massive peaks at $z=15.5$ the fraction of gas available for star
formation in the NoBH control simulation is comparable to the value
found when X-rays are present.  Qualitatively similar conclusions were
reached by Machacek \etal (2003), who showed that an early X-ray
background cannot overcome the negative feedback from \HH\
photodissociation by the soft UV radiation spectrum of the first
stellar sources. Note that, at a distance of 50 comoving kpc, the
X-ray flux from our miniquasars is comparable to the strongest X-ray
background adopted by Machacek \etal (their $\epsilon_{\rm x}=10$
case).  Here we find that, even in the absence of a strong LW flux,
the radiative feedback from X-rays is subtle. It enhances gas cooling
in lower-$\sigma$ peaks that are far away from the initial site of
star formation, thus decreasing the clustering bias of the early
pregalactic population, but does not dramatically reverse or promote
the collapse of pregalactic clouds as a whole.

It has been suggested recently by Oh \& Haiman (2003) that an early
X-ray background would establish an entropy floor over the entire IGM,
thus preventing gas contraction, \HH\ formation, cooling, and the
build up of dense cores in minihaloes. The implication is a large
reduction in the collapsed gas fraction and a pause in the cosmic star
formation history, before more massive haloes with $\Tvir>10^4\,$K
(which can undergo atomic cooling) start forming. We find little
evidence for preheating suppressing \HH\ formation. None of our
minihaloes exhibits an entropy floor and, as we have shown above, the
amount of cold and dense gas in the centres is actually enhanced in
haloes sufficiently removed from the miniquasar. Oh \& Haiman
considered the evolution of a pre-heated, isolated, uniform density
gas parcel with initially primordial \HH\ fraction, and showed that
the entropy floor prevented subsequent \HH\ formation. In our
simulations, however, the heating source is also a catalyst of
\HH. This causes the \HH\ cooling time to become shorter than the
Hubble expansion time at $\delta>40$ (see Fig.~\ref{fig:tcool}), and
thus gas entropy is no longer a conserved quantity inside and in the
vicinity of minihaloes.

\begin{figure*}
\begin{center}
\includegraphics[width=0.48\textwidth]{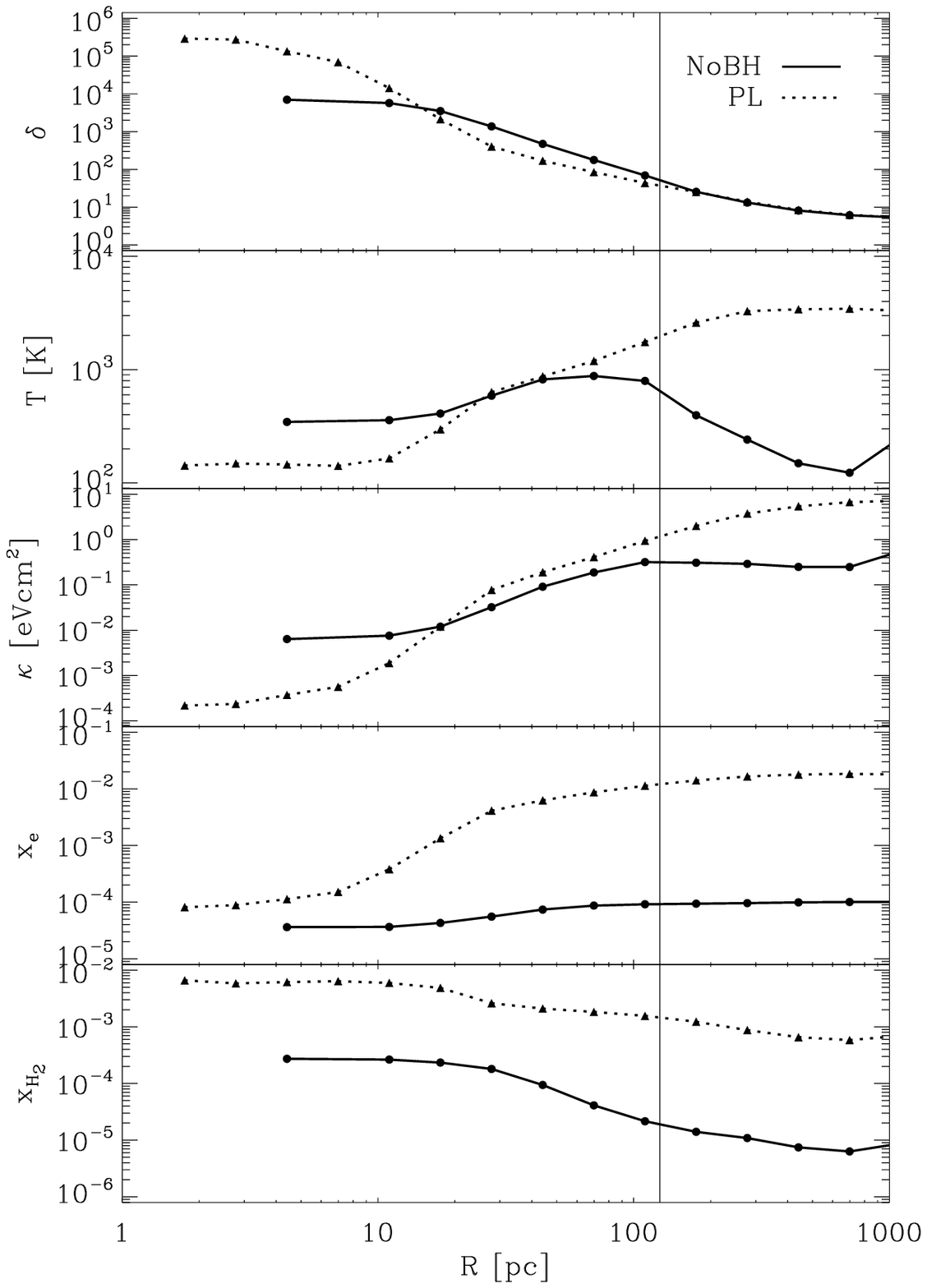}
\includegraphics[width=0.48\textwidth]{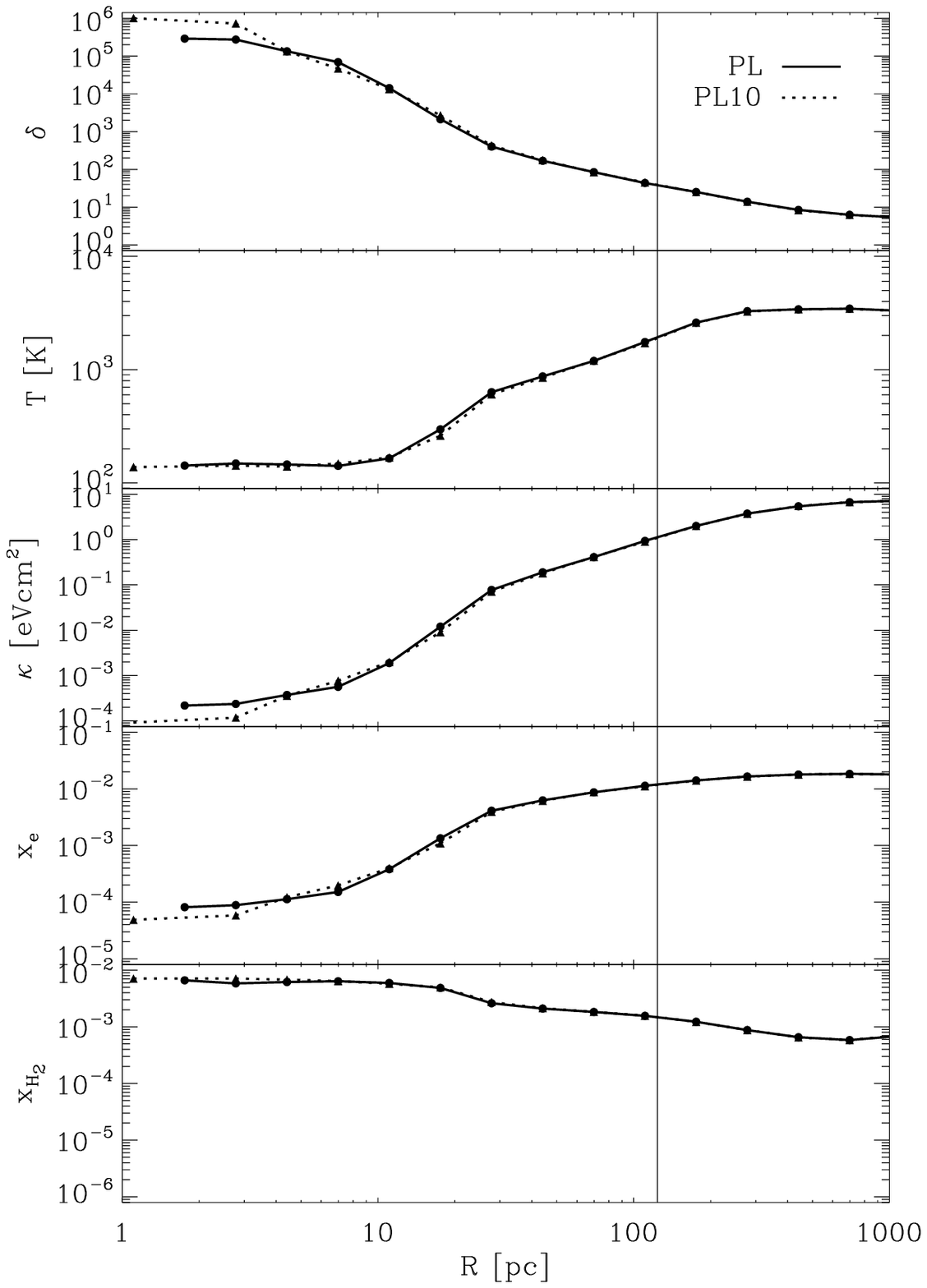}
\end{center}
\caption{
Spherically averaged mass-weighted radial profiles of $\delta$, $T$,
$\kappa$, $x_e$, and $x_{\rm H_2}$ of a representative halo at
$z=17.5$. In the left panel we compare the NoBH with the PL case, in
the right panel the PL with the PL10 case. This halo has a DM mass of
$3\times10^5 \Msun$ and is located 200 comoving kpc from the
miniquasar. See text for discussion.
}
\label{fig:profiles}
\end{figure*}

Fig.~\ref{fig:profiles} shows spherically averaged mass-weighted
profiles of $\delta$, $T$, $\kappa$, $x_e$ and $x_{\rm H_2}$ at $z=17.5$
of a typical low mass minihalo in our simulations. It has a dark
matter mass of $3\times10^5 \Msun$ ($\Tvir \approx 925$ K), a virial
radius of 127 pc (proper), and is located 200 comoving kpc from the
miniquasar. In the NoBH simulation it has a total gas mass of
$5.7\times10^4 \Msun$, a cold gas mass of $1.1\times10^4 \Msun$, and
no cold+dense gas. The central regions of this halo have only
triggered refinement down to level 7, one shy of the maximum
refinement level. It has a central core, where the density profile
levels off at $\delta=7000$. The gas temperature rises from about 100
K to $\Tvir$ at the virial radius, and then drops down to the central
value of $350$ K. This drop is possible due to the slightly elevated
central \HH\ fraction of $2.7 \times 10^{-4}$, which, however, is not
high enough to allow the gas to condense to the cold+dense threshold
density of $330\;{\rm cm}^{-3}$ in a Hubble time. The X-ray flux from
the miniquasar changes this picture dramatically. The \HH\ catalysis
has led to an enhancement of almost 2 orders of magnitude in $x_{\rm H_2}$
outside the virial radius, from $10^{-5}$ to $8\times10^{-4}$. Towards
the centre the increase in temperature and density allows the \HH\
abundance to grow further, peaking at a central value of $x_{\rm H_2}=6.6
\times 10^{-3}$. This increase in $x_{\rm H_2}$ causes the gas to cool
down to 140 K and greatly increases the central overdensity
($\delta=3\times 10^5$), triggering refinement all the way to the
maximum refinement level. In the PL simulation the external medium is
at a temperature of $\gtrsim 3000 $ K, which is much higher than
$\Tvir$. Thus no virial shock developes, and \HH\ cooling causes the
temperature profile to decreases monotonically towards the centre,
dropping below $\Tvir$ at $\sim 40$ pc, a third of the virial
radius. Gas that is hotter than $\Tvir$ cannot accrete on to the halo,
and as a consequence the outer halo regions show a reduced gas
density. The total gas mass decreases by 30 per cent down to $4.0 \times 10^4
\Msun$. The increased cooling, however, has boosted the amount of cold
gas by 60 per cent, up to $1.8 \times 10^4 \Msun$, and even allowed $6.5
\times 10^3 \Msun$ to reach the cold+dense threshold. The entropy
profile reflects the additional cooling in the centre: it also
decreases monotonically and reaches a minimum of $2.2 \times
10^{-4}\;{\rm eV\;cm^2}$, a factor of 30 below the central value in
the NoBH case ($6.4 \times 10^{-3}\;{\rm eV\;cm^2}$).

The spatial resolution of our AMR simulations is determined by the
maximum refinement level, which is usually set by computational
costs. Mesh cells that have refined down to the maximum refinement
level can pose a problem. The physics of cooling and collapsing gas is
inherently unstable: Jeans unstable cells should refine further, but
the simulation does not allow it. This leads to an error in the
numerical solution of the differential equations governing the
system. One way to prevent the gas from continuing to trigger
refinment is to introduce Artificial Pressure Support (APS) to
stabilize the density peaks that have reached maximal refinement
against further collapse. With APS the internal gas energy of every
mesh cell that reaches maximum refinement and is Jeans unstable is
raised to 10 times the value that would stabilize it against
gravitational collapse. This provides support against further
refinement, but comes at the cost of introducing an artificially high
central pressure and temperature inside haloes that have reached
maximum refinement. Neither applying APS nor letting the simulation
progress without it is correctly modelling the physics of the
problem. The only solution is to continue to refine further (Abel
\etal 2002) or to introduce sink particles (Krumholz, McKee, \& Klein
2004). In our fiducial simulations we have 50,000-60,000 mesh cells at
the maximum refinement level of 8. In order to determine whether the
associated numerical error affects the conclusions of our study, we
have re-run the PL simulation from $z=21$ to $z=17.5$, with a maximum
refinement level of 10 (for a maximum dynamic range of 131,072) and
APS. In the right panel of Fig.~\ref{fig:profiles} we plot for the PL
and new 'PL10' runs a comparison of the five profiles described
above. As expected the two profiles agree excellently at radii outside
of the central core ($>10$ pc). Even inside the core, the temperature
and $x_{\rm H_2}$ profiles agree very well. The gas in the PL10 halo,
however, has been able to reach an overdensity a factor of $\sim 3.5$
higher than in the PL case, and has $\sim 50$ per cent more cold+dense
gas. Although we show here only profiles for one halo, we see similar
trends for all haloes that reach maximum refinement, and so we
conclude that the amount of cold+dense gas in haloes in the PL
simulation should be viewed as a lower limit. The numerical error due
to evolving the simulation with cells at the maximum refinement level
is unlikely to affect the global conclusions of this paper.

\section{Discussion}

Active galactic nuclei powered by supermassive holes keep the Universe
ionized at $z\lesssim 4$, structure the IGM, and probably regulate
star formation in their host galaxies. Their seeds were likely planted
at very early epochs through the collapse of massive
stars. Intermediate-mass holes accreting gas from the surrounding
medium may shine as miniquasars at redshifts as high as $z\sim 20$. In
this paper we have carried out AMR cosmological simulations using
\textsc{enzo} to address the thermodynamic effect of miniquasars on
the IGM at early times.

X-ray radiation from the miniquasar efficiently heats the gas in the
simulated box to a volume-averaged temperature of 2800 K after one
Salpeter time ($z=17.5$), and 6150 K after two ($z=15.5$). The main
effect of this sharp increase in temperature is a reduction of gas
clumping in the IGM by as much as a factor of 3, due to the Jeans
smoothing of sheets and filaments in the `cosmic web' by increased
gas pressure. This smoothing will lower the number of hydrogen
recombinations and thus the number of UV photons per baryon required
to reionize the universe. Since X-rays are more efficient at heating
than at ionizing, the free electron fraction is never raised to more
than $\sim 10$ per cent, reaching a volume average of 3 per cent at
$z=17.5$. Provided that either collisions or Ly$\alpha$ radiation can
couple the spin temperature to the kinetic temperature of the largely
neutral gas, it may be possible to observe the X-ray heated region in
redshifted 21-cm line emission against the CMB (Madau, Meiksin, \&
Rees 1997) with future facilities like the {\it LOw Frequency ARray}
({\it LOFAR}).

The elevated free electron fraction leads to a strong enhancement of
molecular hydrogen, both inside haloes and in the more tenuous
filaments. The volume-averaged \HH\ fraction is raised to $x_{\rm H_2} = 4
\times 10^{-5}$, 20 times larger than the primordial value. This will
delay the buildup of a uniform UV photodissociating background by
subsequent sources of Lyman-Werner photons. We plan to carry out a
detailed calculation of this suppression and its consequences in
future work. The increased \HH\ abundance allows gas with
overdensities $\delta>40$ to cool within a Hubble time. As a
consequence the evolution of gas within haloes and filaments is not
adiabatic, and no entropy floor is established. While global heating
suppresses baryonic infall and lowers the gas mass fraction at
overdensities $\delta$ in the range 20-2000, enhanced molecular
cooling increases the amount of dense material at $\delta>2000$. Thus,
while the vast majority of the baryons, both in volume and mass, is
heated up, the densest gas at the centres of haloes is actually cooled
down. The largest relative increase in the amount of cold material
occurs in the mass range $2\times 10^5\,<M_{\rm halo}<10^6\,\Msun$,
where the boosting effect can exceed 1-2 orders of magnitude. Yet for
many haloes in the proximity of our miniquasar ($<75$ kpc) the
\HH-boosting effect is too weak to overcome heating, and the cold and
dense gas mass actually decreases. Overall, the radiative feedback
from X-rays enhances gas cooling in lower-$\sigma$ peaks that are far
away from the initial site of star formation, thus decreasing the
clustering bias of the early pregalactic population, but does not
appear to dramatically reverse or promote the collapse of pregalactic
clouds as a whole.

While we have included the essential physics necessary to model the
radiative feedback from the first miniquasars, our study, like any,
has its limitations. In the following we briefly address some of
these:

\begin{enumerate}

\item We have not considered the environmental impact of the
progenitor of the black hole. In the case of direct collapse of a
primordial gas cloud to an intermediate-mass black hole the issue of a
progenitor star would not arise. The more typical formation mechanism
of the black hole, however, may be the collapse of a Population III
star. In this case the copious UV radiation emitted during its
main-sequence lifetime could have significantly altered the initial
conditions for our simulation. Recent three-dimensional simulations
(O'Shea \etal 2005b) seem to indicate that the heating and ionizing
effect from the Population III progenitor are rather short lived. In
their simulation the gas quickly recombines and cools, \HH\ reforms,
and gas is able to collapse to form a second generation star in a
neighboring halo only 265 pc from the Population III host halo.

\item The strong increase in \HH\ abundance catalyzed by the X-rays is
dependent on the absence of a Lyman-Werner \HH-dissociating
background. As shown by Machacek \etal (2003) the positive feedback
effects from an X-ray background in the presence of a LW background
are rather mild, and they did not find a global increase in \HH\
abundance by a factor of 20. For this reason our work should be
considered as a simulation of the \textit{first} miniquasar in its
volume, \ie before other sources establish a LW background. How soon
after the miniquasar a LW background can be established, given the
large increase in \HH, remains to be seen.

\item Recently there has been a renewed interest in the importance of
deuterium hydride (HD) to the cooling of primordial gas clouds
(Johnson \& Bromm 2005; Lipovka, Nunez-Lopez, \& Avila-Reese 2005;
Nagakura \& Omukai 2005). HD becomes more efficient than \HH\ cooling
at $T<200$K, and, given a sufficient HD abundance, can cool gas down
to below 100 K. In our simulations deuterium chemistry has been
neglected. An increased free-electron fraction, as produced by our
miniquasar, would lead to a catalysis of HD molecules, so it is quite
possible that HD cooling may become important in the cores of our
minihaloes.

\item As the emission spectrum of a miniquasar is uncertain, so are
the photoheating and photoionization rates. We have mainly considered
a pure power-law spectrum, ranging from 0.2 to 10 keV, with a slope of
$-1$. The power-law component in our MCD simulation is slightly softer
($\alpha=-1.2$), but the differences between PL and MCD are
negligible.

Note also that we have neglected UV radiation with $13.6\,
{\rm eV} < h\nu < 200\, {\rm eV}$. If a significant amount of the emitted 
Lyman-continuum flux escapes the host halo, it will be
necessary to follow the radiative transfer in order to fully assess
the environmental influence of the first miniquasars. 

\item We have not taken into account the possibility of a feedback
effect from the miniquasar's radiation on the black hole's mass
accretion rate. Instead of assuming a constant exponential growth
rate, it would be better to determine $\dot{M}$ as a function of time
from the amount of cold gas available for accretion. Unfortunately our
simulations don't resolve the length scale of the accretion flow, and a
self-consistent treatment of the mass accretion rate will have to wait
for higher resolution studies.

\end{enumerate}

\section*{Acknowledgments}

We thank T. Abel, G. Bryan, Z. Haiman, and B. O'Shea for many
informative discussions, Ryan Montgomery for help with the phase
diagrams, and the referee, Marie Machacek, for helpful comments that
improved the flow of the paper. Fig.~\ref{fig:3Dvol} was created with
Nick Gnedin's \textit{\textbf{I}onization \textbf{Fr}ont
\textbf{I}nteractive \textbf{T}ool} (IFrIT). Support for this work was
provided by NSF grants PHY99-07949 and AST02-05738, and by NASA grants
NAG5-11513 and NNG04GK85G (P.M.).  P.M. acknowledges support from the
Alexander von Humboldt Foundation.  M.K. thanks the Graduate
Fellowship Program at Kavli Institute for Theoretical Physics, Santa
Barbara, where part of this work was done. All computations were
performed on NASA's Project Columbia supercomputer system and on
UpsAnd, a Beowulf cluster at UCSC. Movies of the simulations are
available at http://www.ucolick.org/$\sim$mqk/miniqso.

\label{lastpage}


\begin{thebibliography}{}

\bibitem{aazn} Abel, T., Anninos, P., Zhang, Y., Norman, M. L., 1997, New 
Astron., 2, 181 

\bibitem{abn} Abel, T., Bryan, G., Norman, M. L., 2000, ApJ, 540, 39

\bibitem{abn2} Abel, T., Bryan, G., Norman, M. L., 2002, Science, 295, 93

\bibitem{ann} Anninos, P., Zhang, Y., Abel, T., Norman, M. L., 1997, New Astron., 2, 209

\bibitem{bl} Barkana, R., Loeb, A., 1999, ApJ, 523, 54

\bibitem{bac} Bond, J. R., Arnett, W. D., Carr, B. J., 1984, ApJ, 280, 825

\bibitem{bcl2} Bromm, V., Coppi, P. S., Larson, R. B., 2002, ApJ, 564, 23

\bibitem{bl03} Bromm, V., Loeb A., 2003, ApJ, 596, 34

\bibitem{bl04} Bromm, V., Larson, R. B., 2004, ARA\&A, 42, 79

\bibitem{c03} Cen, R., 2003, ApJ, 591, 12 

\bibitem{cfa} Ciardi, B., Ferrara, A., Abel, T., 2000, ApJ, 533, 594

\bibitem{eh} Eisenstein, D. J., Hu, W., 1999, ApJ, 511, 5

\bibitem{ehut} Eisenstein, D. J., Hut, P., 1998, ApJ, 498, 137

\bibitem{fan} Fan, X., et al., 2003, ApJ, 125, 1649  

\bibitem{fwh}  Fryer, C. L., Woosley, S. E., Heger, A., 2001, ApJ, 550, 372

\bibitem{fc}  Fuller, T. M., Couchman, H. M. P., 2000, ApJ, 544, 6

\bibitem{gp}  Galli, D., Palla, F., 1998, A\&A, 335, 403

\bibitem{gb1} Glover, S. C. O., Brand, P. W. J. L., 2001, MNRAS, 321, 385

\bibitem{gb3}  Glover, S. C. O., Brand, P. W. J. L., 2003, MNRAS, 340, 210

\bibitem{hai} Haiman, Z., 2004, ApJ, 613, 36

\bibitem{ham} Haiman, Z., Abel, T., Madau, P., 2001, ApJ, 551, 599

\bibitem{har} Haiman, Z., Abel, T., Rees, M. J., 2000, ApJ, 534, 11

\bibitem{htl} Haiman, Z., Thoul, A. A., Loeb, A., 1996, ApJ, 464, 523

\bibitem{hrl} Haiman, Z., Rees, M. J., Loeb, A., 1997, ApJ, 476, 458

\bibitem{jb} Johnson, J. L., Bromm, V., 2005, MNRAS, submitted (astro-ph/0505304)

\bibitem{kbd} Koushiappas, S. M., Bullock, J. S., Dekel, A., 2004, MNRAS, 354, 292 

\bibitem{kmk} Krumholz, M. R., McKee, C. F., Klein R. I., 2004, ApJ, 611, 399

\bibitem{ls} Lepp, S., Shull, J. M., 1983, ApJ, 270, 578

\bibitem{lnlav} Lipovka, A., Nunez-Lopez, R., Avila-Reese, V., 2005, MNRAS, 361, 850

\bibitem{mba1} Machacek, M. M., Bryan, G. L., Abel, T., 2001, MNRAS, 548, 509

\bibitem{mba3} Machacek, M. M., Bryan, G. L., Abel, T., 2003, MNRAS, 338, 273

\bibitem{mmr} Madau, P., Meiksin, A., Rees, M. J., 1997, 475, 429

\bibitem{mr} Madau, P., Rees, M. J., 2001, ApJ, 551, L27

\bibitem{mad04} Madau, P., Rees, M. J., Volonteri, M., Haardt, F., Oh, S. P.,
2004, ApJ, 606, 484

\bibitem{mak} Makishima, K., \etal, 2000, ApJ, 535, 632

\bibitem{mcol} Miller, M. C., Colbert, E. J. M., 2004, IJMPD, 13, 1

\bibitem{no} Nagakura T., Omukai, K., 2005, MNRAS, submitted (astro-ph/0505599)

\bibitem{ohha} Oh, S. P., Haiman, Z., 2003, MNRAS, 346, 456

\bibitem{onshn} O'Shea, B. W., Nagamine, K., Springel, V., Hernquist, L., Norman, 
M. L., 2005a, ApJ, in press (astro-ph/0312651)  

\bibitem{oawn} O'Shea, B. W., Abel T., Whalen D., Norman M. L., 2005b, ApJ, 
628, L5

\bibitem{reed} Reed, D. S., Bower, R., Frenk, C. S., Gao, L., Jenkins, 
A., Theuns, T., White, S. D. M., 2005, MNRAS, submitted (astro-ph/0504038)

\bibitem{rgs} Ricotti, M., Gnedin, N. Y., Shull, J. M., 2001, ApJ, 560, 580

\bibitem{ric} Ricotti, M., Ostriker, J. P., Gnedin, N. Y., 2005, MNRAS, 357, 207

\bibitem{scha} Schaerer, D., 2002, A\&A, 382, 28

\bibitem{shapa} Shapiro, S. L., 2004a, in Carnegie Observatories 
Astrophysics Series, Vol 1: Coevolution of Black Holes and Galaxies, ed. L. C.
Ho (Cambridge: Cambridge Univ. Press), p.103

\bibitem{shapb} Shapiro, S. L., 2004b, ApJ, 610, 913

\bibitem{sir} Shapiro, P. R., Iliev, I. T., Raga, A. C.,  2004, MNRAS, 348, 753 

\bibitem{svs} Shull, J. M., Van Steenberg, M. E., 1985, ApJ, 298, 268

\bibitem{teg} Tegmark, M., Silk, J., Rees, M. J., Blanchard, A., Abel, T., Palla, 
F., 1997, ApJ, 474, 1 

\bibitem{vgs} Venkatesan, A., Giroux, M. L., Shull, M. J., 2001, ApJ, 563, 1

\bibitem{vr} Volonteri, M., Rees, M. J., 2005, ApJ, in press (astro-ph/0506040)

\bibitem{wan} Whalen, D., Abel, T., Norman, M. L., 2004, ApJ, 610, 22

\bibitem{yahs} Yoshida, N., Abel, T., Hernquist, L., Sugiyama, N., 2003, ApJ, 592, 645

\end{thebibliography}
\end{document}